\documentclass[twocolumn,showpacs,preprintnumbers,amsmath,amssymb,floatfix,nofootinbib,superscriptaddress]{revtex4-1}
\usepackage{amsthm}
\usepackage{graphicx}
\usepackage{dcolumn,xcolor}
\usepackage{bm}
\usepackage[normalem]{ulem}

\newcommand{\be}{\begin{equation}}
\newcommand{\ee}{\end{equation}}
\newcommand{\ba}{\begin{eqnarray}}
\newcommand{\ea}{\end{eqnarray}}
\newcommand{\ban}{\begin{eqnarray*}}
\newcommand{\ean}{\end{eqnarray*}}
\newcommand{\non}{\nonumber}
\newcommand{\eq}[1]{(\ref{#1})}
\newcommand{\n}[1]{\label{#1}}

\newcommand{\hhh}{\hspace{0.5cm}}
\newcommand{\ind}[1]{\mbox{\tiny #1}}
\newcommand{\bi}{\bibitem}

\begin{document}

\title{Conformal Killing horizons and their thermodynamics}
\author{Alex B. Nielsen}
\email{alex.nielsen@aei.mpg.de}
\affiliation{Albert-Einstein-Institut, Max-Planck-Institut f\"ur Gravitationsphysik, D-30167, Hannover, Germany}
\author{Andrey A. Shoom}
\email{ashoom@mun.ca}
\affiliation{Department of Mathematics and Statistics, Memorial University, St. John's, Newfoundland and Labrador, A1C 5S7, Canada}

\begin{abstract}
Certain dynamical black hole solutions can be mapped to static spacetimes by conformal metric transformations.
This mapping provides a physical link between the conformal Killing horizon of the dynamical black hole and a Killing horizon of the static spacetime.
We show how this conformal relation can be used to derive thermodynamic properties for the dynamical black holes.
Although these horizons are defined quasi-locally and can be located by local experiments, they are distinct from other popular notions of quasi-local horizons such as apparent horizons.
Thus in the dynamical Vaidya spacetime describing constant accretion of null dust, the conformal Killing horizon, which is null by construction, is the natural horizon to describe the black hole.
\end{abstract}


\maketitle

\section{Introduction}

The detection of gravitational waves from merging binary black holes \cite{TheLIGOScientific:2016pea} marks the first observations of truly dynamical black holes in our universe. However, even in standard general relativity, there is still no fully satisfactory notion of a dynamical black hole \cite{Ashtekar:2013dza}. The notion of an event horizon depends on the future evolution of the spacetime and its asymptotic structure. It is by definition unobservable to any observer, even one that falls into the black hole \cite{Nielsen:2008cr}. The notion of an apparent horizon, while quasi-local and in principle observable, depends on a choice of spacetime foliation and is thus non-unique. Attempts have been made to address this ambiguity \cite{Bengtsson:2010tj,Hayward:2009ji,Nielsen:2010wq}, but a conclusive answer remains elusive in all but the most symmetrical spacetimes.

As observational techniques improve, the examination of spacetime structure near black hole horizons comes into the realm of observational possibility. Gravitational waves already provide one such technique and very long baseline interferometry provides another \cite{Johannsen:2016vqy}. In this way, a number of theories that postulate physical structure at, or nearby black hole horizons become observationally testable \cite{Giddings:2017jts}. As the sophistication of theoretical models also improves, the issue of where such structure should form, if it forms at all, is of central importance. The existence of quasi-local or even explicitly local conditions that dictate the existence of a horizon are a pre-requisite of any such locally causal model. Recently, the possibility of using curvature invariants to locate horizons has been raised and important theorems governing their existence have been proven in stationary spacetimes \cite{Page:2015aia}.

An important feature widely believed to be associated with black holes is their thermodynamic properties. For black hole thermodynamics an unambiguous identification of the relevant horizon appears essential. The area of the horizon is associated with black hole entropy and broad features of Hawking radiation are determined by evaluating functions on the black hole horizon \cite{FN}. In this thermodynamic sense, the black hole horizon is closely related to its thermodynamic properties. In static spacetimes, the consensus opinion is that the relevant horizon is a Killing horizon of the associated static Killing vector field. For static spacetimes a number of techniques exist to calculate the emission of Hawking radiation or calculate the entropy. Here we will take the relatively standard position that if a static Killing horizon exists, then it is indeed the correct surface to use when calculating thermodynamic properties.

In the dynamical case, however, the issue of which surface correctly defines the black hole and its thermodynamic properties remains open. In fact, even the issue of how to calculate the correct thermodynamic properties for dynamical black holes remains open. For example there is still ambiguity about how the classical surface gravity should be calculated and how this is related to any Hawking emission spectra \cite{Nielsen:2007ac}. It is to this issue that we turn our attention here.

A special class of dynamical spacetimes exist which admit a conformal Killing vector field. For these spacetimes there exists a conformal transformation that can map the spacetime to a static spacetime. In the case where the static spacetime admits the structure of a Killing horizon, the corresponding structure in the dynamical spacetime is the conformal Killing horizon. This conformal Killing horizon is by definition a null surface in the dynamical spacetime and cannot coincide with the \textit{dynamical horizon} of \cite{Ashtekar:2004cn} which is by definition spacelike. In fact, it is not expected that the conformal Killing horizon will coincide with any of the popular quasi-local horizon definitions based on trapped surfaces, such as trapping horizons or apparent horizons, which are typically spacelike when their area is increasing.

Dynamical spacetimes admitting conformal Killing horizons are of particular interest because the existence of this conformal mapping provides a mechanism to translate properties of the dynamical spacetimes to static spacetimes. Standard techniques can then be used to derive the relevant physical observables in the static spacetime. In fact, following an argument of Dicke \cite{Dicke:1961gz}, later extended by other authors \cite{Flanagan:2004bz,Faraoni:2006fx,Deruelle:2010ht}, and extended to the semi-classical regime \cite{Codello:2012sn}, the outcomes of physical experiments are \textit{equivalently} mapped to each other, provided due care is taken over how the conformal transformation affects the coupling of the geometrical spacetime and the matter degrees of freedom that constitute experimental apparatus. This physical equivalence provides the necessary structure for investigating dynamical spacetimes with conformal Killing horizons.

As an explicit example of this equivalence, we examine here the calculation of the Hawking radiation. It has been noted by a number of authors that the standard calculations of the Hawking effect enjoy a conformal covariance, either in purely static cases \cite{Marques:2011uq} or even in dynamical spacetimes \cite{Jacobson:1993pf,Nielsen:2012xu}. While a static black hole has a constant temperature, the conformally related dynamical black hole is expected to have a changing temperature, but whose form is entirely determined by the conformal factor and the static black hole temperature. To take advantage of these conformal relations the relevant horizon in the dynamical case is required to be the null conformal Killing horizon, rather than any dynamical horizon, trapping horizon or apparent horizon. The correct thermodynamic surface to define a dynamical black hole is the conformal Killing horizon and not the apparent horizon.

We show here how this conformal mapping can be used to extract thermodynamic information about the dynamical black hole. For this purpose it is important to note that neither the static, nor dynamical spacetimes are strictly required to be solutions of the Einstein equations with well-motivated stress-energy tensors, nor are the spacetimes required to be isomorphic to one another as is the case in conformally invariant spacetimes. Typically if one spacetime is a solution of the Einstein equations with a particular matter content, then conformally transformed spacetimes will be solutions for conformally transformed matter where the conformal factor will manifest itself as a scalar field.

We use the conventions adopted in \cite{MTW} and geometrized units $G=c=1$.

\section{Basic definitions}

Let ${\cal M}$ be a smooth, $D-$dimensional Lorentzian manifold ($D\geq4$) covered by a set of local coordinate neighbourhoods with coordinates $(x^a, a=1,...,D)$ and endowed with the metric $g_{ab}$. Let $\phi_{\lambda}:{\cal M}\to{\cal M}$ be a one-parameter group of isometries. Its generator, 
\be\n{1.1}
\xi^a=\frac{dx^a(\lambda)}{d\lambda}\,,\hhh x^a(\lambda)=\phi_\lambda(x^a)\,,
\ee 
is called a {\em Killing vector field} on ${\cal M}$. By definition, it satisfies the {\em Killing equation}
\be\n{1.2}
{\cal L}_{\xi}g_{ab}=\nabla_{a}\xi_{b}+\nabla_{b}\xi_{a}=0\,.
\ee
Here $\xi_{a}=g_{ab}\xi^b$ and the derivative operator $\nabla_{a}$ is associated with the metric $g_{ab}$. In what follows, we shall consider a stationary spacetime which has a Killing horizon associated with $\xi^a$ \cite{Carter,FN}. Just outside the Killing horizon $\xi^a$ is timelike, i.e., $\xi^{a}\xi_{a}<0$. 

A {\em Killing horizon} is a null hypersurface ${\cal H}$, which is a smooth co-dimension one embedded submanifold of ${\cal M}$ such that the Killing vector field $\xi^a$ is orthogonal to ${\cal H}$. This orthogonality condition can be written in the following Frobenius form:
\be\n{1.3}
\xi_{[a}\nabla_{b}\xi_{c]}\stackrel{\mathrm{\cal H}}{=}0\,.
\ee
Here and in what follows $\stackrel{\mathrm{\cal H}}{=}$ denotes that the expression is calculated on ${\cal H}$. The orthogonality condition implies that $\xi^a$ is null on ${\cal H}$ and the integral curves of $\xi^{a}$, suitably parametrized, are null geodesic generators of ${\cal H}$. 

Consider now a metric $\bar{g}_{ab}$ conformally related to $g_{ab}$,
\be\n{1.4}
\bar{g}_{ab}=\Omega^2g_{ab}\,,\hhh \bar{g}^{ab}=\Omega^{-2}g^{ab}\,,\hhh \Omega\ne0\,.
\ee
Here $\Omega^2$ is the conformal factor which is a smooth function on ${\cal M}$. For this metric we have
\be\n{1.5}
{\cal L}_{\xi}\bar{g}_{ab}=\left({\cal L}_{\xi}\Omega^2\right)g_{ab}\,.
\ee
This expression implies that if the conformal factor is constant along the Killing orbits, i.e. if ${\cal L}_{\xi}\Omega^2=0$, then $\xi^{a}$ is a Killing vector field for $\bar{g}_{ab}$. Otherwise, $\xi^{a}$ is called a {\em conformal Killing vector field} for $\bar{g}_{ab}$ and it satisfies the {\em conformal Killing equation}
\be\n{1.6}
{\cal L}_{\xi}\bar{g}_{ab}=\bar{\nabla}_{a}\bar{\xi}_{b}+\bar{\nabla}_{b}\bar{\xi}_{a}=\left(\ln\Omega^2\right)_{,c}\xi^{c}\,\bar{g}_{ab}=\frac{2}{D}(\bar{\nabla}_{c}\xi^{c})\bar{g}_{ab}\,.
\ee
Here
\be\n{1.7}
\bar{\xi}_{a}=\bar{g}_{ab}\xi^b=\Omega^2\xi_{a}\,
\ee
and the derivative operator $\bar{\nabla}_{a}$ is associated with the metric $\bar{g}_{ab}$. 
 
In analogy with a Killing horizon, one can define a conformal Killing horizon \cite{DH, SultanaDyer}. A {\em conformal Killing horizon} is a null hypersurface $\bar{\cal H}$, which is a smooth co-dimension one embedded submanifold of ${\cal M}$ such that the Killing vector field $\xi^a$ is orthogonal to $\bar{\cal H}$.

Thus, we see that the conformal transformation \eq{1.4} mathematically maps a Killing horizon ${\cal H}$ corresponding to the Killing vector field $\xi^{a}$ for $g_{ab}$ to a conformal Killing horizon $\bar{\cal H}$ corresponding to the conformal Killing vector field $\xi^{a}$ for $\bar g_{ab}$. 

\section{An example in linear Vaidya spacetime}

Let us now consider an example of conformally static, ingoing Vaidya space-time with $D=4$ illustrating the location of a conformal Killing horizon. The spacetime metric reads \cite{Vaidya:1951zz,Hiscock:1982pa,Alex} 
\ba\n{3.1}
d\bar{s}^2&=&-\left(1-\frac{2m(v)}{r}\right)dv^2+2dvdr+r^2d\omega^2\,,\\
d\omega^2&=&d\theta^2+\sin^2\theta d\phi^2\,.\non
\ea
Any choice of $m(v)$ will give a solution of the Einstein equations sourced by null dust, but here we consider the following linear mass function:
\be\n{3.3}
m(v)=M+\mu (v-v_{0})\,,
\ee
where $M\geq0$ and $v_{0}$ are constants and $\mu = dm/dv$ plays the role of a mass flux parameter, vanishing in the static limit.\footnote{Approximate values for the parameter $\mu$ in cases of astrophysical relevance are given in \cite{Nielsen:2010gm}. For example, for a solar mass black hole accreting at the Eddington rate, $\mu$ would take a dimensionless value on the order of $10^{-21}$. In the limit $\mu=0$ this describes a Schwarzschild solution of mass $M$.} The full causal structure of this spacetime is discussed in \cite{Kuroda:1984pp,Hiscock:1982pa}. Here we take the quasi-local approach of assuming this metric only in some quasi-local region where our experiments are to be performed.\footnote{For definiteness we could consider attaching this spacetime to a Schwarzschild spacetime of mass $M$ in the past, before $v_{0}$, and in the future, after a time $v_{1}$, with a mass $M+\mu(v_{1}-v_{0})$, but this is not necessary for our purposes here.} This Vaidya space-time has a conformal Killing vector 
field
\be\n{3.4}
\xi^{a}=\frac{m(v)}{M}\delta^a_v+\frac{\mu r}{M}\delta^a_r\, ,
\ee
satisfying the conformal Killing equation \eq{1.6}. Note that there is a freedom to rescale this conformal Killing field by a constant $c$: $\xi^{a} \rightarrow c\,\xi^{a}$. The normalization chosen here ensures that in the static limit, $\mu \rightarrow 0$, $\xi^{a}$ becomes a Killing vector field of a Schwarzschild spacetime with unit norm at infinity. This conformal Killing vector field is null at the conformal Killing horizon given by
\be\n{3.5}
\xi^{a}\xi_{a} =\frac{m(v)}{M^2}\left[2\mu r-m(v)\left(1-\frac{2m(v)}{r}\right)\right] = 0.
\ee
For the above form of $m(v)$, Eq. \eq{3.5} has solutions at
\be\n{3.6}
r=\frac{m(v)}{4\mu}\left(1\pm\sqrt{1-16\mu}\right)\,.
\ee
This equation has real solutions if $\mu \leq 1/16$ and these two solutions define two conformal Killing horizons. The negative sign choice in Eq. \eq{3.6} maps smoothly to the standard Killing horizon in the static limit $\mu\rightarrow 0$ and the positive sign choice becomes infinitely large in the Schwarzschild limit. In addition to the two conformal Killing horizons, there is also a single spherically symmetric trapping horizon, which occurs at $r=2m(v)$, inside the inner conformal Killing horizon. This coincides with the negative sign choice in Eq. \eq{3.6} in the static limit. Henceforth, in what follows, we concentrate on this negative sign choice. Note that one can locate the conformal Killing horizon of the Vaidya space-time by using the scalar polynomial curvature invariants of \cite{McNutt:2017gjg}.

The conformal factor needed to transform the metric (\ref{3.1}) into a static spacetime, using equation (\ref{1.4}), is
\be\n{3.10}
\Omega^2 = \frac{2M^2}{m(v)r}\, .
\ee
In this case the vector field (\ref{3.4}) will become a static Killing vector field of the metric $\bar{g}_{ab}$. The coordinate $r$ will not be the areal radius coordinate and the metric depends explicitly on the coordinate $v$. Again, this conformal factor has a normalization freedom under rescaling by a constant and the normalization choice here is chosen to have dimensionless conformal factor and a finite non-vanishing result in the limit $\mu\rightarrow 0$, so that $M$ is the local mass. The resulting spacetime is not asymptotically flat as $\Omega$ is not unity at infinity. In fact there are two Killing horizons of the resultant spacetime and the Killing vector field (\ref{3.4}) is only timelike between them.

\section{Thermodynamic properties}

On the conformal Killing horizon the conformal Killing vector coincides with the outgoing radial null vector field. In this linear Vaidya case, the change of the mass function along the horizon is proportional to the same mass function
\be
{\cal L}_{\xi}m = \frac{\mu m}{M}
\ee
and the change in the horizon area is proportional to the area
\be
{\cal L}_{\xi}A = \frac{2 \mu A}{M}\, .
\ee
Both of these vanish in the static limit of $\mu \rightarrow 0$ as expected, a result due to the choice of finite normalization in \eq{3.4}. Since the mass function represents the energy of the black hole and one quarter of the area represents the entropy in the standard picture of black hole thermodynamics, an effective thermodynamic temperature can be obtained by taking the ratio of these two.
\be
T_{\ind{eff}} = \frac{4{\cal L}_{\xi}m}{{\cal L}_{\xi}A} = \frac{2m}{A} = \frac{4\mu^2}{\pi m(v)(1-8\mu - \sqrt{1-16\mu})}.
\ee
In the static limit, $\mu\rightarrow 0$ this ratio is finite, and gives the expected temperature of a Schwarzschild black hole, $T=1/8\pi M$. This is a purely formal result. A more operational quantity related to the Hawking radiation can be obtained from the geometric surface gravity, $\kappa$, and its relation to the Hawking temperature $T = \kappa/2\pi$.

It has long been known that the geometric surface gravity of a black hole can be defined in a conformally invariant way \cite{Jacobson:1993pf}. This conformal invariance holds also for the dynamical spacetime case considered here. A straightforward calculation of the $\kappa_{1}$, defined as $\nabla_a(\xi^b\xi_b)=-2\kappa_1\xi_a$ (see Eq. (1) of \cite{Jacobson:1993pf}), yields
\be
\kappa_{1} = \frac{2\mu\sqrt{1-16\mu}}{M(1-\sqrt{1-16\mu})}\, .
\ee
This value is constant, which may at first be surprising for a dynamical spacetime, but recall that this value is invariant under conformal transformations and therefore takes the same value in the static spacetime related by the conformal factor (\ref{3.10}). In the limit $\mu \rightarrow 0$, it takes the value $1/4M$ which is the expected Schwarzschild value. This factor is determined by the normalization of \eq{3.4}. The standard value is for measurements made at infinity, although it is possible to calculate the surface gravity at other locations using the redshift factor \cite{Wald:1999vt}. In the static limit, to obtain this measurable temperature, the Killing vector should be normalized to coincide with the four-velocity of the observer at the observer's location via the freedom $\xi^{a}\rightarrow{\tilde{\xi}}^a =c\,\xi^{a}$ with $c$ a constant. 

For linear Vaidya, the normalization can be performed in the static spacetime related to the Vaidya spacetime by the conformal factor \eq{3.10}, giving
\be
{\tilde{\xi}}^{a} = c\,\xi^{a} = \frac{1}{\sqrt{\left(\frac{2m(v)}{r}\left(1-\frac{2m(v)}{r}\right)-4\mu\right)}}\xi^{a}\, .
\ee
with this prefactor evaluated at the location of the particular observer.

Using this normalization allows us to compute a geometric surface gravity in the static spacetime for that particular observer. However, one further feature needs to be considered. Since we want the static spacetime to encode exactly the same physics as the linear Vaidya spacetime, we require by the arguments of Dicke \cite{Dicke:1961gz} that the coupling of geometry to matter be non-trivial in the static spacetime. The geometric surface gravity will not be the actual measured surface gravity since the measured surface gravity will be a factor $1/\Omega$ times this value. This scaling ensures that the physical predictions of the two conformal frames are equivalent \cite{Flanagan:2004bz}. The surface gravity for an observer following the trajectory of the conformal Killing vector in the dynamical Vaidya spacetime will be given by $\kappa = \Omega c \kappa_{1}$ which gives
\be
\kappa = \frac{r^{1/2}}{\sqrt{m^2 r - 2m^3 -2\mu mr^2}}\frac{2\mu\sqrt{1-16\mu}}{(1-\sqrt{1-16\mu})}\, ,
\ee
where this should be evaluated at the location of the observer. The evaluation of this dynamical black hole surface gravity is our main result. As can be checked by inspection, this surface gravity is decreasing as the observer moves along her/his trajectory and thus the measured Hawking temperature will decrease as the black hole increases in mass, in line with intuitive expectations.

Strictly speaking this result is only valid in the region of spacetime described fully by the Vaidya metric with mass function given by (\ref{3.3}). Since the Hawking effect is not an intrinsic property of a region of spacetime but depends also on boundary conditions, this result also depends on boundary conditions and these need to be treated with some care when the observer is beyond the region described by the Vaidya metric or particle production is occurring outside this region. The static spacetime that we have used for this result is rather unusual in that it is not a solution of the Einstein equations, but the conformally transformed Einstein equations. Its static Killing horizon has an increasing Wald entropy \cite{Wald:1993nt}, although constant area. The geometric surface gravity is constant although the measured Hawking temperature is decreasing.

\section{Discussion}

We have argued that in the case of conformally static spacetimes, the correct surface to identify the black hole and its thermodynamic properties is the conformal Killing horizon. Although quasi-local, this surface is null by definition and does not coincide with the dynamical horizon or apparent horizon. For a black hole that eventually changes its mass accretion rate, this conformal Killing horizon will not coincide with the event horizon either. We have calculated an explicit example of the linear Vaidya spacetime and shown how a time-varying temperature can be derived. This result uses two main ingredients. Firstly that in a static spacetime, the correct horizon for determining quasi-local physics is given by the standard Killing horizon \cite{Wald:1999vt} and secondly that the physical predictions for conformally related spacetimes are equivalent \cite{Flanagan:2004bz}. Together these two concepts allow a calculation of the Hawking effect in a dynamical black hole spacetime.

The class of linear Vaidya spacetimes thus provides a test bed for ideas about dynamical black holes, where there is sufficient geometrical structure to guide us from known static results. The relevant horizon surface is neither a null event horizon nor a quasi-local apparent horizon, although it retains features of both. It is worth mentioning that the idea that the horizon should be null even in dynamical spacetimes is compatible with discussions in \cite{Wall:2009wm} that argue the correct boundary for a definition of the generalized second law should be a null boundary.

\begin{acknowledgments}

A. A. S. is grateful to the Natural Sciences and Engineering Research Council of Canada Discovery Grant No. 261429-2013 for its financial support.

\end{acknowledgments}


\begin{thebibliography}{99}

\bibitem{TheLIGOScientific:2016pea}
  B.~P.~Abbott {\it et al.} [LIGO Scientific and Virgo Collaborations],
  Phys.\ Rev.\ X {\bf 6} (2016) no.4,  041015
  doi:10.1103/PhysRevX.6.041015
  [arXiv:1606.04856 [gr-qc]].

\bibitem{Ashtekar:2013dza}
  A.~Ashtekar,
  Gen.\ Rel.\ Grav.\  {\bf 46} (2014) 1706
  doi:10.1007/s10714-014-1706-2
  [arXiv:1312.6425 [gr-qc]].

\bibitem{Nielsen:2008cr}
  A.~B.~Nielsen,
  Gen.\ Rel.\ Grav.\  {\bf 41} (2009) 1539
  doi:10.1007/s10714-008-0739-9
  [arXiv:0809.3850 [hep-th]].

\bibitem{Bengtsson:2010tj}
  I.~Bengtsson and J.~M.~M.~Senovilla,
  Phys.\ Rev.\ D {\bf 83} (2011) 044012
  doi:10.1103/PhysRevD.83.044012
  [arXiv:1009.0225 [gr-qc]].

\bibitem{Hayward:2009ji}
  S.~A.~Hayward,
  Phys.\ Rev.\ D {\bf 81} (2010) 024037
  doi:10.1103/PhysRevD.81.024037
  [arXiv:0905.3950 [gr-qc]].  

\bibitem{Nielsen:2010wq}
  A.~B.~Nielsen, M.~Jasiulek, B.~Krishnan and E.~Schnetter,
  Phys.\ Rev.\ D {\bf 83} (2011) 124022
  doi:10.1103/PhysRevD.83.124022
  [arXiv:1007.2990 [gr-qc]].  

\bibitem{Johannsen:2016vqy}
  T.~Johannsen, C.~Wang, A.~E.~Broderick, S.~S.~Doeleman, V.~L.~Fish, A.~Loeb and D.~Psaltis,
  Phys.\ Rev.\ Lett.\  {\bf 117} (2016) no.9,  091101
  doi:10.1103/PhysRevLett.117.091101
  [arXiv:1608.03593 [astro-ph.HE]].

\bibitem{Giddings:2017jts}
  S.~B.~Giddings,
  Nature Astronomy 1, Article number: 0067 (2017)
  doi:10.1038/s41550-017-0067
  [arXiv:1703.03387 [gr-qc]].

\bibitem{Page:2015aia}
  D.~N.~Page and A.~A.~Shoom,
  Phys.\ Rev.\ Lett.\  {\bf 114} (2015) no.14,  141102
  doi:10.1103/PhysRevLett.114.141102
  [arXiv:1501.03510 [gr-qc]].
  
\bi{FN} V. P. Frolov and I. D. Novikov, {\em Black Hole Physics: Basic Concepts and New Developments}, (Kluwer Academic Publishers, Dordrecht, The Netherlands, 1998).

\bibitem{Nielsen:2007ac}
  A.~B.~Nielsen and J.~H.~Yoon,
  Class.\ Quant.\ Grav.\  {\bf 25} (2008) 085010
  doi:10.1088/0264-9381/25/8/085010
  [arXiv:0711.1445 [gr-qc]].
  
\bibitem{Ashtekar:2004cn}
  A.~Ashtekar and B.~Krishnan,
  Living Rev.\ Rel.\  {\bf 7} (2004) 10
  doi:10.12942/lrr-2004-10
  [gr-qc/0407042].
  
\bibitem{Dicke:1961gz}
  R.~H.~Dicke,
  Phys.\ Rev.\  {\bf 125} (1962) 2163.
  doi:10.1103/PhysRev.125.2163 

\bibitem{Flanagan:2004bz}
  E.~E.~Flanagan,
  Class.\ Quant.\ Grav.\  {\bf 21} (2004) 3817
  doi:10.1088/0264-9381/21/15/N02
  [gr-qc/0403063]. 
 
\bibitem{Faraoni:2006fx}
  V.~Faraoni and S.~Nadeau,
  Phys.\ Rev.\ D {\bf 75} (2007) 023501
  doi:10.1103/PhysRevD.75.023501
  [gr-qc/0612075].

\bibitem{Deruelle:2010ht}
  N.~Deruelle and M.~Sasaki,
  Springer Proc.\ Phys.\  {\bf 137} (2011) 247
  doi:10.1007/978-3-642-19760-4\textunderscore 23
  [arXiv:1007.3563 [gr-qc]].

\bibitem{Codello:2012sn}
  A.~Codello, G.~D'Odorico, C.~Pagani and R.~Percacci,
  Class.\ Quant.\ Grav.\  {\bf 30} (2013) 115015
  doi:10.1088/0264-9381/30/11/115015
  [arXiv:1210.3284 [hep-th]].
 
\bibitem{Marques:2011uq}
  G.~T.~Marques and M.~E.~Rodrigues,
  Eur.\ Phys.\ J.\ C {\bf 72} (2012) 1891
  doi:10.1140/epjc/s10052-012-1891-7
  [arXiv:1110.0079 [gr-qc]]. 
  
\bibitem{Jacobson:1993pf}
  T.~Jacobson and G.~Kang,
  Class.\ Quant.\ Grav.\  {\bf 10} (1993) L201
  doi:10.1088/0264-9381/10/11/002
  [gr-qc/9307002]. 
  
\bibitem{Nielsen:2012xu}
  A.~B.~Nielsen and J.~T.~Firouzjaee,
  Gen.\ Rel.\ Grav.\  {\bf 45} (2013) 1815
  doi:10.1007/s10714-013-1560-7
  [arXiv:1207.0064 [gr-qc]].  
  

\bi{MTW} C. W. Misner, K. S. Thorne, and J. A. Wheeler, {\em Gravitation}, (W. H. Freeman and Co., San Francisco, 1973).

\bi{Carter} B. Carter, in {\em Black Holes: Les Houches 1972}, eds. C. DeWitt and B. S. DeWitt (Gordon and Breach Science Publishers, Inc. New York, N.Y., 1973).



\bi{DH} C.~C.~Dyer and E.~Honig, 
J.\ Math.\ Phys.\ {\bf 20}, 409 (1979)
doi: http://dx.doi.org/10.1063/1.524078.

\bi{SultanaDyer} J.~Sultana and C.~C.~Dyer,
J.\ Math.\ Phys.\ {\bf 45}, 4764 (2004)
doi: http://dx.doi.org/10.1063/1.1814417

\bibitem{Vaidya:1951zz}
  P.~Vaidya,
  Proc.\ Natl.\ Inst.\ Sci.\ India A {\bf 33} (1951) 264.

\bibitem{Hiscock:1982pa}
  W.~A.~Hiscock, L.~G.~Williams and D.~M.~Eardley,
  Phys.\ Rev.\ D {\bf 26} (1982) 751.
  doi:10.1103/PhysRevD.26.751

\bi{Alex} A. B. Nielsen,
Galaxies 2014,{\bf 2}, 62 
doi:10.3390/galaxies2010062

\bibitem{Nielsen:2010gm}
  A.~B.~Nielsen,
  Class.\ Quant.\ Grav.\  {\bf 27} (2010) 245016
  doi:10.1088/0264-9381/27/24/245016
  [arXiv:1006.2448 [gr-qc]].

\bibitem{Kuroda:1984pp}
  Y.~Kuroda,
  Prog.\ Theor.\ Phys.\  {\bf 72} (1984) 63. 
 
\bibitem{McNutt:2017gjg}
  D.~D.~McNutt and D.~N.~Page,
  Phys.\ Rev.\ D {\bf 95} (2017) no.8,  084044
  doi:10.1103/PhysRevD.95.084044
  [arXiv:1704.02461 [gr-qc]].
 
\bibitem{Wald:1993nt}
  R.~M.~Wald,
  Phys.\ Rev.\ D {\bf 48} (1993) no.8,  R3427
  doi:10.1103/PhysRevD.48.R3427
  [gr-qc/9307038].

\bibitem{Wald:1999vt}
  R.~M.~Wald,
  Living Rev.\ Rel.\  {\bf 4} (2001) 6
  doi:10.12942/lrr-2001-6
  [gr-qc/9912119].

\bibitem{Wall:2009wm}
  A.~C.~Wall,
  JHEP {\bf 0906} (2009) 021
  doi:10.1088/1126-6708/2009/06/021
  [arXiv:0901.3865 [gr-qc]].

\end{thebibliography}
\end{document}